\documentclass[12pt]{article} 
\usepackage{a4,amssymb,amsmath}
\begin{document} 
\newcommand{\eqs}{\begin{eqnarray*}} 
\newcommand{\ens}{\end{eqnarray*}} 
 
\newcommand{\eqa}{\begin{eqnarray}} 
\newcommand{\ena}{\end{eqnarray}} 
\newcommand{\eq}{\begin{equation}} 
\newcommand{\en}{\end{equation}} 
 
\newtheorem{theorem}{Theorem}[section] 
\newtheorem{lemma}[theorem]{Lemma} 
\newtheorem{corollary}[theorem]{Corollary} 
\newtheorem{proposition}[theorem]{Proposition} 
\newtheorem{example}[theorem]{Example}

\font\smrm=cmr9
\font\smbf=cmbx9
\font\smit=cmmi9

\font\fett=cmmib10
\font\fetts=cmmib7
\font\bigbf=cmbx10 scaled \magstep2
\font\bigrm=cmr10 scaled \magstep2
\font\bbigrm=cmr10 scaled \magstep3
\def\cnl{\centerline}
\def\etal{{\it et al.}}
\def\text{{}}
\def\un{^{(n)}}
\def\ff{{\cal F}}
\def\Ref#1{(\ref{#1})}

\def\Blm{\left|}
\def\Brm{\right|}
\def\Bl{\left(}
\def\Br{\right)}
\def\nti{n\to\infty}
\def\lnti{\lim_{\nti}}
\def\law{{\cal L}}
\def\sji{\sum_{j\ge1}}
\def\Cal{\cal}

\def\real{\text{\rm I\kern-2pt R}}
\def\re{\real}
\def\expec{\text{\rm I\kern-2pt E}}
\def\ex{\expec}
\def\prob{\text{\rm I\kern-2pt P}}
\def\pr{\prob}
\def\rat{\text{\rm Q\kern-5.5pt\vrule height7pt depth-1pt\kern4.5pt}}
\def\comp{\text{\rm C\kern-4.7pt\vrule height7pt depth-1pt\kern4.5pt}}
\def\nat{\text{\rm I\kern-2pt N}}
\def\integ{{\bf Z}}
\def\qedbox{\vcenter{\hrule height.5pt\hbox{\vrule width.5pt height8pt
\kern8pt\vrule width.5pt}\hrule height.5pt}}
\def\half{{\textstyle {1 \over 2}}}
\def\quarter{{\textstyle {1 \over 4}}}
\def\threequarters{{\textstyle {3 \over 4}}}
\def\Blb{\left\{}
\def\Brb{\right\}}

\def\sjmi{\sum_{j=1}^{m_i}}
\def\sjmij{\sum_{j=1}^{m_{il}}}

\def\scrl{{\Cal L}}
\def\scrd{{\Cal D}}
\def\tod{\buildrel \scrd \over \longrightarrow}
\def\eqd{\buildrel \scrd \over =}
\def\tolone{\buildrel L_1 \over \longrightarrow}
\def\toltwo{\buildrel L_2 \over \longrightarrow}
\def\tolp{\buildrel L_p \over \longrightarrow}
\def\l{\lambda}
\def\L{\Lambda}
\def\a{\alpha}
\def\b{\beta}
\def\g{\gamma}
\def\G{\Gamma}
\def\d{\delta}
\def\D{\Delta}
\def\e{\varepsilon}
\def\h{\eta}
\def\z{\zeta}
\def\th{\theta}
\def\k{\kappa}
\def\m{\mu}
\def\n{\nu}
\def\p{\pi}
\def\r{\rho}
\def\s{\sigma}

\def\t{\tau}
\def\f{\varphi}
\def\ch{\chi}
\def\ps{\psi}

\def\lee{\,\le\,}
\def\leee{\quad\le\quad}
\def\gee{\,\ge\,}
\def\geee{\quad\ge\quad}
\def\scrn{{\Cal N}}
\def\scra{{\Cal A}}
\def\scrf{{\Cal F}}
\def\var{\text{\rm Var\,}}
\def\cov{\text{\rm Cov\,}}

\def\sin{\sum_{i=1}^n}
\def\sn{\sum_{i=1}^n}
\def\sln{\sum_{l=1}^n}
\def\cross{\times}

\def\nin{\noindent}
\def\bsk{\bigskip}
\def\msk{\medskip}
\def\widebar{\bar}
\def\ignore#1{}
\def\adb#1{{\bf #1}}
\def\adba#1{{\sl #1}}

\def\uo{^{(0)}}
\def\ui{^{(1)}}
\def\ut{^{(2)}}
\def\uh{^{(3)}}
\def\uj{^{(j)}}
\def\uk{^{(k)}}
\def\uI{^{(i)}}
\def\ul{^{(l)}}
\def\Corr{{\rm Corr\,}}
\def\OU{{\rm OU\,}}
\def\hb{{\hat\b}}
\def\hs{{\hat\s}}
\def\bs{\bar \s}
\def\ts{\tilde\s}
\def\hl{{\hat\l}}
\def\hbui{\hb\ui}
\def\hbuh{\hb\uh}
\def\bbui{\bar\b\ui}
\def\tbui{\tilde\b\ui}
\def\bbuh{\bar\b\uh}
\def\tbuh{\tilde\b\uh}
\def\non{\nonumber}

\title{On inter-species regression analysis}

\author{Brigitte Pallmann, A.~D.~Barbour,\thanks{Angewandte Mathematik, 
Winterthurerstrasse~190, CH--8057 Z\"urich, Switzerland} \and 
D.~J.~Hosken,\thanks{Centre for Ecology and Conservation,
University of Exeter in Cornwall, Penryn,
TR10 9EZ, UK} 
~and P.~I.~Ward\thanks{Zoologisches Museum, Winterthurerstrasse 190, 
CH--8057 Z\"URICH, Switzerland}
\\
Universit\"at Z\"urich}
\date{March 10th, 2006}
\maketitle

\nin{\bf Running head:} Inter-species regression analysis

\bsk\bsk
\nin{\bf Corresponding author:}  A. D. Barbour \hfil\break
 A.D.Barbour@math.unizh.ch,
Tel: +41 44 635 5846,\ \ Fax: +41 44 635 5705.

\ignore{
\begin{abstract}
When conducting inter-species
regression analyses, the phylogenetic relationships between
the individual species need to be taken into account.  In this
paper, a procedure for conducting such analyses is discussed,
which only requires the use of a measure of relationship between
pairs of species, rather than a complete phylogeny, and 
which at the same time assesses the importance to be attached to 
the relationships with regard to the conclusions reached.
The procedure is applied to data from Minder~{\it et al\/}.~(2005),
relating testis size to
mean hind tibia length, duct length and spermathecal area in
15 species of Scathophagidae (Diptera).
We show that considering the phylogenetic structure significantly
improves the fit of the model to the data.
We find a robust relationship between 
testis size and spermathecal area but could not support one between 
testis size and spermathecal duct length.
\end{abstract}
}

\ignore{
\title[On inter-species regression analysis]{On inter-species regression analysis}
\author[Pallmann]{Brigitte Pallmann}
\address[Brigitte Pallmann]{Angewandte 
Mathematik, Universit\"at Z\"urich, Winterthurerstrasse 190, CH-8057, Z\"urich, 
Switzerland.}
\author[Barbour]{A. D. Barbour}
\address[A. D. Barbour]{Angewandte 
Mathematik, Universit\"at Z\"urich, Winterthurerstrasse 190, CH-8057, Z\"urich, 
Switzerland. 
}
\email{a.d.barbour@math.unizh.ch}
\author[Hosken]{D. J. Hosken}
\address[D. J. Hosken]{Zoologisches Museum, Winterthurerstrasse 190, 
CH--8057 Z\"URICH, Switzerland.
}
\author[Ward]{P. I. Ward}
\address[P. I. Ward]{Zoologisches Museum, Winterthurerstrasse 190, 
CH--8057 Z\"URICH, Switzerland.
}
\email{pward@zoolmus.unizh.ch}
\date{April 11, 2005}
\thanks{ADB was supported in part by Schweizer\-ischer 
Nationalfonds\-projekte Nrs 20-67909.02 and 20-107935/1,
and by the Institute for Mathematical Sciences of the National
University of Singapore. DJH was supported 
in part by Schweizerischer Nationalfonds Projekt Nr.\ 31-56902.99. 
}
}

\ignore{
\leftskip=1cm
\rightskip=1cm
\nin{\smbf Abstract.} {\smrm When conducting inter--species
regression analyses, the phylogenetic relationships between
the individual species need to be taken into account.  In this
paper, a procedure for conducting such analyses is discussed,
which only requires the use of a measure of relationship between
pairs of species, rather than a complete phylogeny, and 
which at the same time assesses the importance to be attached to 
the relationships with regard to the conclusions reached.
The procedure is applied to data from Minder~{\smit et al\/}.~(2005),
relating testis size to
mean hind tibia length, duct length and spermathecal area in
15 species of Scathophagidae.
We show that considering the phylogenetic structure significantly
improves the fit of the model to the data.
We find a robust relationship between 
testis size and spermathecal area but could not support one between 
testis size and spermathecal duct length.}
\leftskip=0cm
\rightskip=0cm
}

\newpage
\begin{abstract}
When conducting inter-species
regression analyses, the phylogenetic relationships between
the individual species need to be taken into account.  In this
paper, a procedure for conducting such analyses is discussed,
which only requires the use of a measure of relationship between
pairs of species, rather than a complete phylogeny, and 
which at the same time assesses the importance to be attached to 
the relationships with regard to the conclusions reached.
The procedure is applied to data from Minder~{\it et al\/}.~(2005),
relating testis size to
mean hind tibia length, duct length and spermathecal area in
15 species of Scathophagidae (Diptera).
We show that considering the phylogenetic structure significantly
improves the fit of the model to the data.
We find a robust relationship between 
testis size and spermathecal area but could not support one between 
testis size and spermathecal duct length.
\end{abstract}

\bsk\bsk
\nin {\bf Keywords:} likelihood based inference, Ornstein-Uhlenbeck process, 
phylogenetic relationship.

\newpage

 
\section{Introduction}\label{intro} 
 \setcounter{equation}{0} 

Comparative studies are a widely employed and powerful tool in 
evolutionary investigations.  They have been used to elucidate 
macro--evolutionary patterns for many phenomena, including testis and 
sperm size evolution (e.g.\ Gage 1994; Hosken 1997), brain size 
evolution (e.g.\ Martin 1981; Pagel \& Harvey 1989) and the scaling of 
metabolic rates (e.g.\ Thompson \& Withers 1998; McNab 2002).  Formerly, 
species values of characters of interest were regressed against 
putative predictor variables to elucidate possible relationships 
(e.g.\ Cummins \& Woodall 1985).  However, because of common descent, 
species do not represent independent data points, and hence species 
level analyses based on simple regression analyses have been 
criticised (Harvey \& Pagel 1991).  This is 
not to say that phylogeny has primacy of cause over other factors, 
merely that species level analyses may be misleading (Harvey~2000). 
For example, a simulation study by Martins \& Garland~(1991) 
investigated the across-species association between two variables, and 
found that the Type~I error rate was~16\%; when they 
employed phylogenetic control, the error in the regression was 
reduced to~5\%.  This paper presents a procedure for
conducting regression analyses in the presence of phylogenetic
relationships, which also assesses the importance
of these relationships in the analysis.  The procedure is based
on the Ornstein--Uhlenbeck model of the way in which 
inter-species differences
evolve, and is closely related to the simplest version 
of Hansen's~(1997) approach.

In order to derive our procedure, we begin with a more
detailed exposition of the underlying problem.
In the classical regression model, the value~$y$ of the
`dependent' variable of interest is expressed linearly in 
terms of the values $x\ui,x\ut,\ldots$ of a number of 
explanatory `covariables', up to an additive `error'~$e$,
which accounts for any variation in~$y$ not attributable
to the covariables.  Thus, for each of~$n$ observations
indexed by~$i$, $1\le i\le n$, we write
\eq\label{(1)} 
y_i = \b\uo + \b\ui x_i\ui + \b\ut x_i\ut +\cdots+\b\uk 
  x_i\uk + e_i,  
\en 
where the~$\b\uj$, $0\le j\le k$, are the coefficients
which relate the values of the covariables to that of~$y$,
and the~$e_i$ are needed because, in practice, it is
usually impossible to find values $\b\uo,\b\ui,\ldots,\b\uk$
such that
\eq\label{(2)} 
y_i = \b\uo + \b\ui x_i\ui + \b\ut x_i\ut +\cdots+\b\uk 
  x_i\uk  
\en 
is {\it exactly\/} true for all~$i$, if $n\ge k+2$.  The values of
the parameters~$\b\uj$ are estimated and their significance
tested with reference to some probabilistic model, which is
assumed to govern the values~$e_i$ of the errors actually
occurring; the simplest assumption is that the~$e_i$,
$1\le i\le n$, arise as realizations of {\it independent\/}
random variables $\e_i$, $1\le i\le n$, which have zero mean and
common unknown variance~$\s^2$, and are normally distributed.

If the indices $1\le i\le n$ in fact represent~$n$ species, 
as in the setting introduced above, and if the measured 
values~$y_i$ and $x_i\ui,\ldots,x_i\uk$ are `typical' values
for the species,
the assumption of independent errors may well be
violated.  This is because the variation
in~$y$ unexplained by the~$x\uj$ can be thought of in part as
resulting from evolutionary change in other, unobserved
explanatory covariables, so that closely related species can
be expected to exhibit rather similar values of~$e$.  Hence,
when conducting regression analyses with such data, it seems
important to take the phylogeny into account (Harvey \& Pagel, 1991).

The reasons for doing so are quite simple, and have long been
understood in the context of quite general regression models. 
When the errors~$\e_i$ in such a model are in fact correlated,
 ordinary least
squares (OLS) procedures still give parameter estimates which
have the right expectation and are asymptotically consistent.  
However, as is especially relevant
when the number of species is fixed and perhaps not large,
their precision is less than that of the best estimates
possible for the actual correlation structure [Draper \& Smith~(1966: 
 80)]. Moreover, estimates of the precision of the OLS estimates, 
calculated in accordance with OLS assumptions, may be
seriously in error 
[see, for
example, Scheff\'e~(1959: 339--343 and \S 10.4)]. 
In such cases, significance tests based on OLS are dangerous.
In the particular context of
inter--species regression, these features of the OLS analysis
were noted by Pagel~(1993).

There is nonetheless still some debate 
about the efficacy of such phylogenetic control.  This is primarily 
because the covariance of traits is explained by ecology and 
phylogeny, which typically overlap; hence, by controlling for 
phylogeny,  variance due to 
ecology is also removed because of the overlap (McNab 2002). 
Despite such arguments, most investigators today use some form 
of phylogenetic control.

A number of methods have been proposed for incorporating
phylogeny into the regression, including trait mapping (Ridley 1985),
nested analysis of variance (Bell 1989, Stearns 1992), pairwise comparison  
(Felsenstein 1985, M\o ller \& Birkhead 1992), independent contrasts 
(Felsenstein 1985), and more directly Ornstein--Uhlenbeck based
analyses (Hansen \& Martins 1996): see also Lynch~(1991)
and Freckleton, Harvey \& Pagel (2002). These methods all involve 
the use of a pre-existing 
phylogeny. Here, we propose a simple and effective procedure, which
can either be applied using a known phylogeny, or else
just using an inter--species distance matrix from which 
a phylogeny could potentially be constructed.
The procedure has the advantage that it not only
only respects the phylogeny, but also allows one to gauge its
importance in the analysis. It also takes proper account
of the (unknown) value at the root of the phylogeny.

\section{Procedure}\label{Proc}

The basic idea is to return to
the underlying assumption, that the~$e_i$'s result from a process
of evolution along the branches of the phylogenetic tree.  The
evolutionary model that we use, 
which can be thought of as a natural generalization of
that of independent, normally distributed errors with common
variance, supposes that the `error' component evolves along
each branch of the phylogenetic tree as an Ornstein--Uhlenbeck
(O-U) process, as discussed in some detail in Felsenstein~(1988)
and in Hansen \& Martins~(1996).  
The O-U process looks locally in time like a Brownian
motion, as would naturally be the result of many small random
genetic changes; however, it also has a tendency to return towards zero,
which can be thought of as the result of selective pressure
acting against departures from equality in~\Ref{(2)}, the strength
of the tendency being larger for larger departures.  For our 
purposes, the main features of the process are that it is a
time-reversible Markov process, that its values
are normally distributed, and that its autocorrelations decay 
exponentially with elapsed time. It is also important in
acting as a good approximation to a wide variety of processes
that result from the combination of random disturbances with a
tendency to move back towards zero, in much the same way that the
normal distribution is frequently a good approximation to sums of weakly
dependent random variables. It thus represents a plausible null
model for describing the `errors' arising during evolution:
see, for instance, Lande~(1976). 
Its distributions are entirely characterized by the 
diffusion constant (infinitesimal variance)~$\t^2$ of its locally 
Brownian behaviour
and by the exponential decay rate~$\l$ of correlations;
its equilibrium distribution is normal, with mean zero and
variance $\s^2 = \t^2/(2\l)$.
We shall denote such a process by $\OU(\t^2,\l)$.

Our model supposes that an $\OU(\t^2,\l)$ process starts in
equilibrium at the root of the phylogenetic tree, and runs,
with time corresponding to distance along the branch, until
the first split.  At this point, its value is taken as the
initial value for two {\it independent\/} $\OU(\t^2,\l)$
processes, which then continue to run along the two branches
until they split again; and so on.  The species, the leaves
of the trees, are assigned as values of~$e_i$ the values of
the $\OU(\t^2,\l)$--processes at the ends of the final~$n$
branches.  This model of evolution along the tree results
in values~$e_i$ realized from jointly normally distributed
random variables $\e_1,\ldots,\e_n$ having equal 
variances~$\s^2$, but now with correlations
\eq\label{(3)} 
C_{il}(\l) := \Corr(\e_i,\e_l) = e^{-\l d_{il}}, 
\en 
where~$d_{il}$ is the distance between species~$i$ and 
species~$l$ along the tree; see, for example, Hansen \& Martins (1996:
Equation~(7)).  Details are given in the Appendix.

The value of the decay rate~$\l$, in combination with
the values of the~$d_{il}$, is seen from~\Ref{(3)} to determine the
importance of the inter-species correlations in
the analysis.
However, the dependence between the errors is better understood
from the following explicit representation.  If two species
$i$ and~$l$ diverge at a time at which the value of the $\OU(\t^2,\l)$ 
process for their common ancestral species
takes the value~$X_0$, and if the evolutionary distances to $i$ and~$l$
from this time until the present are $d_i$ and~$d_l$, respectively,
then the values of the O-U processes $X\uI$ and~$X\ul$ taken by the 
species $i$ and~$l$ can be written as
\eqa
  X\uI &=& X_0 e^{-\l d_i} + V\uI;\non\\
	X\ul &=& X_0 e^{-\l d_l} + V\ul.\label{(4)}
\ena
Here, $X_0$, $V\uI$ and~$V\ul$ are {\it independent\/}, and the latter
two random variables have zero means and variances
$$
  {\t^2 \over 2\l}(1 - e^{-2\l d_i}) \quad{\rm and}\quad
	{\t^2 \over 2\l}(1 - e^{-2\l d_l}),
$$
respectively.		
The correlation between $X\uI$ and~$X\ul$ arises
solely as a result of the elements containing~$X_0$, the value at their
common ancestral species, and these elements can be seen to decline
at exponential rate~$\l$ as evolutionary distance increases.  In the
extreme in which $\l\to0$, the elements containing~$X_0$ both remain 
constant at the value~$X_0$, and the Brownian diffusion model with
diffusion constant~$\t^2$ results.  In the extreme in which $\l\to\infty$,
the elements containing~$X_0$ both tend to zero, and the values $X\uI$ 
and~$X\ul$ become independent, implying independent errors for each
species.  Thus, when fitting the
$\OU(\t^2,\l)$ model to data, a best fit with large~$\l$
indicates more or less independent species, and one with~$\l$
very small indicates a Brownian-like model of evolution, a
fact noted also by Blomberg \etal~(2003) (with their 
parameter~$d$ corresponding to our~$e^{-\l}$).
Neither limit is, however, entirely free of surprises:
see the Appendix~\S4.2 for more details.

Regression analysis based on this model is easy if the
phylogenetic tree --- specifically, all the tree--distances~$d_{il}$
between pairs of species --- are known.  Then, for any fixed~$\l$,
$0<\l<\infty$, the problem reduces to a generalized least squares
analysis: the correlation matrix~$C(\l)$ can be calculated,
and maximum likelihood for the linear model with normally
distributed errors and known correlation matrix 
can be used to find estimates $\hb\uo(\l),\hb\ui(\l),\ldots,
\hb\uk(\l)$ and~$\hs^2(\l)$ of the remaining model parameters,
together with~$L(\l)$, the maximum value of the log--likelihood for
this value of~$\l$.  The value~$\hl$ to be used as an estimate
of the true value of~$\l$ is now obtained by maximizing~$L(\l)$
iteratively with respect to~$\l$, for instance using a golden
section search. This, as in Hansen~(1997: 1345), yields the 
final parameter estimates
\[ 
\hl,\hb\uo(\hl),\ldots,\hb\uk(\hl),\hs^2(\hl) 
\] 
for the regression.  

This rather simple procedure has an important drawback.  If, for
fixed~$\t^2$, the value of~$\l$ becomes very small, the
variance~$\s^2 = \t^2/(2\l)$ of the equilibrium distribution 
becomes large, and, as a result, any particular observed value has
correspondingly low likelihood.  Indeed, for such~$\l$, all
values are strongly related to the (unknown) value at the root,
which is itself a single value chosen from the equilibrium
distribution.  This leads to a large negative contribution to
the log likelihood,  reflecting nothing more than the
potential variability in the value at the root, whose effects
are clearly visible in Hansen~(1997: Tables 1 and~2). It may
seem unnatural to include this in a comparative study, in which
the value of the overall mean~$\b\uo$ is typically of little
interest.  There is also a companion, biological consideration;
for very small~$\l$, it is doubtful whether enough
time can have elapsed for the value at the root to have reached
statistical equilibrium.  In view of this, we prefer to centre
all the covariables $x\ui,\ldots,x\uk$ at zero, and to base
the analysis on the likelihood derived from the joint
distribution of the {\it centred\/} $y$-values
$$
  y_1 - \bar y,\ldots,y_n - \bar y,
$$
where $\bar y$, as usual, denotes the overall mean of the observations.
Because the covariables have been centred, the model now becomes
\eq\label{(5)}
  y_i - \bar y =  \b\ui x_i\ui + \b\ut x_i\ut +\cdots+\b\uk
  x_i\uk + \tilde\e_i, 
\en
where $\tilde\e_i = \e_i - \bar\e$.  The overall mean~$\b\uo$
no longer appears in the model, all other parameters have
their original meaning, and the log likelihood no longer
converges to~$-\infty$ as $\l\to0$, but instead approaches
that of the Brownian model.  Other attempts to circumvent
this difficulty, used in Blomberg \etal~(2003) and in Butler
\&~King~(2004), are discussed in the Appendix, at the end
of~\S4.1; neither seems to be entirely satisfactory.

For each given~$\l$, the linear model theory also gives the
standard deviations to be associated with the parameter
estimates~$\hb\uj(\l)$, $0\le j\le k$, and these can be used
with $\l=\hl$ as reasonable approximations to the standard
deviations of the estimates $\hb\uj(\hl)$, and hence for
tests of hypotheses.  However, $\hl$ has been chosen from the
data, and this source of variability is not included in such
`plug--in' approximations; simulating data samples from
the model obtained from the estimated parameters, and then
using an identical estimation procedure, gives an alternative 
way of judging the actual precision obtained, as well as indicating
any possible bias.  If the value
of~$\l$ is itself of interest, an approximate 95\% confidence
region based on large sample theory is given by the set of
all~$\l$ such that
\eq\label{(6)}
L(\hl)-L(\l) \le 2  
\en
[c.f. Edwards~(1972: 80), Hansen~(1997: 1345)]. This 
region may include $\l=\infty$, 
in which case an analysis that neglects inter--species correlations 
should still be reliable. Again, simulating data samples
from the estimated model gives another measure of the variability
in the estimates of~$\l$.

In practice, the phylogenetic tree is never known precisely,
complete with distances.  However, the method proposed here
can be expected to give useful results, even when the 
distances~$d_{il}$ are only approximately known; as long as
the correlation structure is reasonably represented, gross
errors in the conclusions arising from this source should be
avoided.  Thus, if any molecular or morphological data for
the species are available, on the basis of which a tree can
be reconstructed, this can be carried out, and the corresponding
tree distances used for the~$d_{il}$.  

Alternatively, this
relatively difficult step can be avoided by using the
morphological and molecular data to define a measure of
distance between pairs of species --- in any case, often the
starting point for a tree construction ---  and by then using
these `raw' distances directly in place of the~$d_{il}$. 
This procedure may seem controversial, but it should not be.
If the
raw distances are rather close to being tree distances, then the
tree constructed from them should yield inter-species
distances which are not very different, and the results of the procedure
will change correspondingly little.  However, if the raw distances are 
not particularly
tree-like,  then the tree constructed from them may well yield
rather different inter--species distances, but without any guarantee that
they result in a more reliable picture of the actual correlations.
Computationally feasible tree growing algorithms provide intelligent 
heuristics, but offer no guarantee of finding the correct
phylogeny.  However, using the raw distances,
one at least has tangible data as input, rather than output from
a black box, and our procedure, because of the freedom to choose
the value of~$\l$, still gives a reasonable
idea of how strongly relationship (expressed in terms of the raw
distances) affects correlation.  The phylogeny may not have been
determined, but the analysis still makes reasonable allowance for 
inter--species similarity.

In theory, there may be a problem if the raw distances are 
too far from being tree distances, because the resulting
matrices~$C(\l)$ need not then be positive semi--definite
for all values of~$\l$, as has to be the case for correlation
matrices.  However, Bochner's theorem [Defant \& Floret (1993: 316)] 
implies that $p$'th powers of $l_2$--distances, for 
$p\le 2$, never give rise to this problem, and that the same
applies if a distance can be represented as a sum of such
distances; thus, for instance, Hamming distance
(number of mismatches) can be used for molecular data, and
can be added to Euclidean distances between morphometrical
characters.

Computer programs, written in~R, for performing both estimation and 
simulation from the estimated model, can be obtained from 
the authors.

\section{Example}\label{Ex} 

The procedure is illustrated by application to data in Minder~\etal~(2005), 
with a regression of testis size~$y$ as a function of
mean hind tibia length (HTL)~$x\ui$, spermathecal duct length~$x\ut$ 
and spermathecal 
area~$x\uh$ in~$15$ species of Scathophagidae (Diptera; true flies) 
[Table~1].  In the paper above,
a corresponding analysis was made using the
comparative analysis by independent contrasts program (CAIC) 
(Purvis \& Rambaut 1994) to
correct for the phylogeny, which was deduced from that of
Bernasconi \etal~(2000), itself derived from inter--specific
differences in the sequence of 810 mDNA letters coding for the COI gene.
Here, we look only at the~$15$ species considered by Minder~\etal~(2005).

We consider three evolutionary distance matrices~$d$.  The first,
$d\ui$, is derived from the phylogeny depicted in Bernasconi \etal~(2000:
Fig.~1, 313), with the distance $d\ui_{il}$ between species
$i$ and~$l$ represented by the level in the tree at which their
phylogenies merge (leaves at level~$0$, nearest neighbours at
distance~$1$, etc.).  This tree was rather carefully constructed
from the COI data, using information about the positions of codons
relative to the reading frame, A--T richness, and so on.  Our second
evolutionary distance matrix~$d\ut$ is much cruder, being based
solely on the numbers of mismatches~$d_{il}\uo$ between the COI
sequences for species $i$ and~$l$ [Table~2]: we set
\eq\label{(7)} 
d_{il}\ut = - \log(1-d_{il}\ui/105). 
\en 
This matrix is chosen merely to reflect the fact that, as with most
evolutionary models, the proportion of mismatches converges to a
limit exponentially fast as evolutionary distance increases.
Here, it is assumed that the limiting proportion of mismatches
is $105/810$, which is probably rather small in the context,
but serves to exaggerate any effect caused by the non-linearity
of the proportion of mismatches as a function of time.
The third matrix~$d\uh$ that we consider is that of the model in which
the errors are independent and identically distributed, so that
$d\uh_{il} = \infty$ for all $i\ne l$; this,
however, can be obtained also as the limit as $\l\to\infty$ of
the two previous models.

The procedure described in Section~2, with the correlation 
matrix~$C(\l)$ calculated
for each given value of~$\l$ by substituting the evolutionary distance 
matrix~$d\ui$ into~\Ref{(3)}, shows that~$x\ut$, spermathecal duct length, has 
no appreciable
influence on~$y$. Leaving out this covariable, the log--likelihood is
maximized at a value of $27.17$, with $\hl = 0.0714$ and
with structural parameter estimates
\eq\label{(8)}
\hbui = 0.2353\quad{\rm and}\quad \hbuh = 24.13, 
\en
and with $\hat\t^2 = 0.00912$. The standard deviations of $\hbui$ 
and~$\hbuh$, as
calculated from $C(\hl)$ and~$\hat\t^2$, are $0.078$ and~$8.05$ respectively,
and their estimated correlation is~$-0.294$ (cf.~Younger (1985:
Sections 11.5--11.8)). The approximate $95$\% confidence interval
for~$\l$ calculated according to~\Ref{(6)} was $[0,0.6]$.

We describe the variability and dependence implied by the estimated
model by first evaluating a quantity MSD, the median over all
pairs of species $i$ and~$l$ of
$$
  {\rm SD\,}(i,l) := \sqrt{\ex\{(\e_i-\e_l)^2\}}
    = {\hat\t\over\sqrt{\hat\l}}\,\sqrt{(1 - e^{-\hat\l d_{il}})}\,,
$$
the standard deviation of the difference between the errors for
species $i$ and~$l$, as calculated for the estimated model
(see~\Ref{A3} below).  MSD
is thus a measure of the typical variability to be expected in such
differences.  We then consider the values of
$$
  {\rm RSD\,}(i,l) :=  {\rm SD\,}(i,l)/{\rm MSD}.
$$
If dependence has little effect on the analysis ($\l$ large),
then all such values are close to~$1$; if a pair $(i,l)$ is strongly
dependent, then RSD$\,(i,l)$ is close to zero.  For the analysis here,  
we consider a pair $(i_1,l_1)$ ({\it Norellia striolata\/} and 
{\it N.\ spinimona\/}) which are very closely related, and another,
$(i_2,l_2)$ ({\it Scathophaga sulla\/} and {\it 
S.\ furcata\/}), which are moderately closely related, as examples.
For the analysis just conducted, we find that
$$
  {\rm MSD} = 0.2711,\quad {\rm RSD\,}(i_1,l_1) = 0.3460 \quad{\rm and}\quad
  {\rm RSD\,}(i_2,l_2) = 0.7002.
$$

The maxima of the log--likelihood for the submodels obtained by 
omitting either $x\ui$ or~$x\uh$ are both substantially
more than~$2$ smaller than that obtained above (differences $3.47$ 
and~$3.24$ respectively), indicating that, at the 5\%~level, neither
submodel should be preferred to the model estimated above.  
This means that hind tibia length~(HTL) and spermathecal 
area are associated with testis size across the Scathophagidae after 
phylogenetic control using~$d\ui$. 
Minder~\etal\ (2005) concluded that testis size was related to 
spermathecal area and spermathecal duct length, but not with HTL, 
after phylogenetic control. Our analysis supports 
the relationship with spermathecal area, making this a robust 
conclusion; especially since Minder~\etal\ (2005) also found it with 
a species comparison. The relationship with HTL is intuitively 
appealing, as the simplest expectation is that, when a species gets 
larger, so do all of its body parts. Since the testis area to 
spermathecal duct length relationship is not 
robust in the different analyses, some caution must be exercised when 
considering the relationship reported in Minder~\etal\ (2005).

In order to judge the validity of the procedure, and to obtain an
alternative assessment of the variability in the estimates, the
model~$(8)$ and the correlation structure~$C(\hl)$ for the errors
were held fixed, and data from this model distribution were simulated
$1'000$ times.  The estimation procedure was then applied individually
to each of the $1'000$ resulting sets of data.  These led to mean
values
\eq\label{(9)}
\bbui = 0.2353,\quad {\rm and}\quad \bbuh = 24.14 
\en
for the structural parameter estimates, with empirical standard deviations of
$0.0855$ and~$8.00$ and an empirical correlation of~$-0.201$ for 
the $1'000$ estimates of $\b\ui$ and~$\b\uh$.  
The mean values $\bbui$ and~$\bbuh$ in~\Ref{(9)} are well in accord
with the regression parameters $\hbui$ and~$\hbuh$ of the model~\Ref{(8)}, 
as are the empirical standard deviations 
of the estimates $\hbui$ and~$\hbuh$ with the
values calculated from $C(\hl)$ and~$\hat\t^2$.

The estimates of variability and dependence in the simulated data were
by no means as stable. 
This is not particularly surprising, in view of the
small number~(15) and large variability (${\rm MSD} = 0.2711$, as compared
with estimated effects $\hbui \times {\rm SD\,}(x\ui) = 0.0912$
and $\hbuh \times {\rm SD\,}(x\uh) = 0.0524$) of the observations. 
In just over $40$\% of the simulations,
$\l$~was estimated to be zero, and in a further~$5$\% to be infinity;
the median value was~$0.05$, close to the actual value $0.0714$,
and the~$90$\% confidence interval 
was  $[0,0.69]$. For the quantity MSD, the empirical mean
over the~$1'000$ simulations was $0.2303$, rather lower than the 
true value of~$0.2711$, and the empirical standard deviation was
$0.0591$. The negative bias in the empirical mean is to be expected, 
just as in the classical
model with independent errors, where maximum likelihood
makes no allowance for the fitted degrees of freedom when estimating the
standard deviation. The empirical standard deviation of the MSD values 
is very much
what would be expected when estimating a standard deviation from
only~$15$ observations. The values of RSD$\,(i_1,l_1)$ had a minimum value
of~$0.2966$, taken when~$\l$ was estimated to be zero, and were
thus heavily skewed; the median was~$0.3288$, not far from the
true value of~$0.3460$, and its $90$th percentile was at~$0.7000$.
Analogously, the values of RSD$\,(i_2,l_2)$ had a minimum of
$0.6567$, a median of $0.7002$, to be compared with the true
value of $0.7223$, and a $90$th percentile at~$0.9833$.  Thus,
despite the wide variability in the estimated values of~$\l$, the
results of the simulations as regards variability and
dependence still showed a reasonable consistency.

The same analyses can also be conducted with correlations
based on the distance matrix~$d\ut$.  The results are
broadly the same; the covariable~$x\ut$ is immediately dropped,
and the model with $x\ui$ and~$x\uh$ has likelihood more than~$2$
larger than that of either of the models with just one covariable.
The parameter estimates in this model are
$$
  \hb\ui = 0.2068 \quad{\rm and}\quad \hb\uh = 24.11,
$$
with estimated standard errors of $0.0929$ and~$8.13$, respectively,
all of which are reasonably consistent with~\Ref{(8)}.  For the estimated
error structure, we have
$$
  {\rm MSD} = 0.2558,\quad {\rm RSD\,}(i_1,l_1) = 0.3947
  \quad{\rm and}\quad {\rm RSD\,}(i_2,l_2) =  0.8056, 
$$
the last two values indicating rather weaker evolutionary dependence 
than that found using~$d\ui$, but not outstandingly so.
The main differences between the results with $d\ui$ and~$d\ut$
are that the maximum of the log likelihood using~$d\ut$
(at $25.57$) is smaller by~$1.6$ than that for~$d\ui$, suggesting
that using the cruder matrix~$d\ut$ leads to a somewhat less
good fit, and that the decision to keep~$x\ui$ in the model
is based on a likelihood difference of~$2.12$, quite a bit smaller 
than that obtained using~$d\ui$, indicating that the less good fit
indeed entails some loss of precision, though not enough to change
any of the main conclusions.  The results of simulations
confirmed the reliability of the procedure using~$d\ut$ in much the 
same way as it did when using~$d\ui$.

If phylogenetic correlation is entirely neglected, and the model
with independent and identically distributed errors is used, a
significant loss of precision is observed.  The variable~$x\ut$
is still immediately rejected, and the structural parameters are
estimated by
$$
  \hb\ui = 0.2065 \quad{\rm and}\quad \hb\uh = 19.82,
$$
fairly much as before; variability is estimated by ${\rm MSD}
= 0.2674$ (here corresponding to a residual standard deviation
of around $0.19$), and all RSD--values are~$1$.  This model,
however, has a log likelihood of only $23.53$, more than~$2$ smaller than
either of the two models fitted using phylogenetic correlation,
and would therefore be rejected in comparison to them.
Furthermore, under independence, the model with just~$x\uh$
has log likelihood $22.42$, only~$1.1$ smaller, corresponding
to a two-sided P-value of about~$14$\%. This might suggest that~$x\ui$
should also be omitted from the model; however, the
alternative for the effect of~$x\ui$ is clear and one-sided, 
and the relevant P-value is more properly about~$7$\%, giving 
weak support for retaining~$x\ui$ in the model.  Nonetheless, if
phylogenetic correlation were neglected, there could be a danger 
that the effect of HTL on testis size would be missed.

\section{Discussion}\label{Disc} 

The method that we propose for conducting inter-species
regression analyses is developed from the engagingly simple idea
of using likelihood-based methods in conjunction with a
stationary
Ornstein-Uhlenbeck model of evolution.  The result is a 
procedure which can be carried out without knowing the complete
phylogeny --- a measure of the evolutionary distance between
pairs of species suffices --- and which, at the same time,
assesses the importance of inter-species relationship for the analysis.
It is thus rather surprising that these advantages have not been
emphasized earlier.

\ignore{
The procedure involving branch length modification in Blomberg
\etal~(2003) seems to come closest.  Their model is one
without covariates, $y_i = \b\uo + e_i$, and their main
interest is therefore in the strength of the correlations
between the~$e_i$; we are principally interested in estimating
and testing regression coefficients, and properly accounting
for the correlations is a means towards this end.  It should be
noted that the analysis given in Blomberg 
\etal~(2003, Appendix~2) is not quite correct, in that they set the
variance of the character at the root of the phylogenetic
tree to be~$0$, but do not then take into account the
fact that the mean of the OU--process progressively returns 
from this value towards~$0$.
Their formulae then also have the drawback of involving not 
only the inter-species
distances, but also the root to tip distances, requiring
a more precise knowledge of the underlying phylogeny
than is the case for our analysis: see the Appendix.
}

Our example illustrates that the method works much as
expected when applied to a `typical' biological data set.
Although the detailed
correlation structure was not reliably estimable, because there
were few data points and a low signal to noise ratio,
this still did not prevent the 
regression coefficients and their precisions being
successfully estimated.  Indeed, the procedure performed
very well in a number of respects. It estimated the regression
parameters and the variability of these estimates satisfactorily;
it highlighted the extent to which the highly variable
data did not support accurate estimation of the underlying
dependence structure; and it indicated that, with these data,
replacing the phylogeny with a crude estimate of the inter-species
distances had little impact on the final conclusions.
\newpage
\nin{\bf Acknowledgement}

\msk
ADB was supported 
in part by Schweizerischer Nationalfonds Projekte Nr.\ 20-67909.02
and 20-107935/1,
and by the Institute for Mathematical Sciences of the National
University of Singapore. DJH was supported 
in part by Schweizerischer Nationalfonds Projekt Nr.\ 31-56902.99.

\vfil\eject 
\section*{References} 

\nin
Bell G. 1989.
A comparative method.
{\it Amer.\ Nat.\/}~{\bf 133}, 553--71. 

\msk\nin
Bernasconi MV, 
Pawlowski J, Valsangiacomo  C, Piffaretti J-C,   
Ward  PI. 2000.
Phylogeny of Scathophagidae (Diptera, Calyptratae) based 
on mitochondrial DNA sequences.  {\it Mol.\ Phyl.\ Evol.\/}~{\bf 16}, 308--15.

\msk\nin
Blomberg SPB, Garland T Jr, Ives AR.  2003.
Testing for phylogenetic signal in comparative data: behavioural traits
are more labile.   {\it Evolution\/}~{\bf 57}, 717--45. 

\msk\nin
Butler MA, King AA. 2004.
Phylogenetic comparative analysis: a modelling approach for adaptive
evolution.  {\it The American Naturalist\/}~{\bf 164}, 683--95.

\msk\nin
Cummins JM, 
Woodall PF. 1985.
On mammalian sperm dimensions. {\it J.~Reprod.\ Fert.\/}~{\bf 75}, 153--75. 

\msk\nin 
Defant A, 
 Floret K. 1993.
{\it Tensor norms and operator ideals.\/}
North Holland Mathematical Studies 176. Amsterdam: North Holland. 

\msk\nin 
Draper NR, 
Smith H. 1966.
{\it Applied regression analysis.\/} New York: Wiley. 

\msk\nin 
Edwards AWF. 1972.
{\it Likelihood.\/} Cambridge University Press. 

\msk\nin
Felsenstein J. 1985. 
Phylogenies and the comparative method.  
{\it Amer.\ Nat.\/}~{\bf 125}, 1--15. 

\msk\nin
Felsenstein J. 1988.
Phylogenies and quantitative characters.
{\it Ann.\ Rev.\ Ecol.\ Syst.\/}~{\bf 19}, 445--71. 

\msk\nin
Freckleton RP, Harvey PH, Pagel M. 2002.  Phylogenetic analysis
and comparative data: a test and review of evidence.
{\it Amer.\ Nat.\/}~{\bf 160}, 712--26.

\msk\nin
Gage MJG. 1994. 
Associations between body size, mating pattern, 
testis size and sperm lengths across butterflies. 
{\it Proc.\ Roy.\ Soc.~B\/}~{\bf 258}, 247--54.

\msk\nin
Hansen TF, Martins EP.  1996.
Translating between microevolutionary process and macroevolutionary
patterns: the correlation structure of interspecific data.
{\it Evolution\/}~{\bf 50}, 1404--17.

\msk\nin
Hansen TF. 1997.
Stabilizing selection and the comparative analysis of adaptation.
{\it Evolution\/}~{\bf 51}, 1341--51.   

\msk\nin
Harvey PH. 2000. 
Why and how phylogenetic relationships should be 
incorporated into studies of scaling.  In: Brown JH, West GB, eds,
 {\it Scaling in Biology,\/} 253--265. Oxford University Press.

\msk\nin
Harvey PH, 
Pagel MD. 1991.
{\it The comparative method in evolutionary biology.\/}
Oxford University Press. 

\msk\nin
Hosken DJ. 1997. 
Sperm competition in bats. {\it Proc.\ Roy.\ Soc.~B\/}~{\bf 264}, 385--92. 

\msk\nin
Lande R. 1976.
Natural selection and random genetic drift in phenotypic evolution.  
{\it Evolution\/}~{\bf 30}, 314--34.

\msk\nin
Lynch M. 1991. Methods for analysis of comparative data in evolutionary
biology.  {\it Evolution\/}~{\bf 45}, 1065--80.

\ignore{\nin
Markow T.\ A., 2002. 
Perspective: Female remating, operational sex ratio, 
and the arena of sexual selection in Drosophila species. 
Evolution~56, 1725-1734 }

\msk\nin
Martin RD. 1981. 
Relative brain size and basal metabolic rate in 
terrestrial vertebrates.  {\it Nature\/}~{\bf 293}, 57--60. 

\msk\nin
Martins EP,
Garland T Jr. 1991.
Phylogenetic analyses of the correlated evolution of continuous characters:
a simulation study.  
{\it Evolution\/}~{\bf 45}, 534--57. 

\msk\nin
Math\'eron G. 1965.
{\it Les variables regionalis\'ees et leur estimation.\/}
Paris: Editions Masson. 

\msk\nin
McNab BK. 2002.  
{\it The physiological ecology of vertebrates: a view from 
energetics.\/} Ithaca, New York:
Comstock Publishing Associates, Cornell University Press.

\msk\nin 
Minder AM, 
Hosken DJ, Ward PI.  2005.
Co-evolution of male and female reproductive characters across the 
Scathophagidae (Diptera).  {\it  J.\ Evol.\ Biol.\/}~{\bf 18}, 60--69. 
 
\ignore{\nin
Minder, A., 2002. 
Co-evolution of sperm, testis and the female tract 
morphology in Scathophagidae. MSc Thesis, University of Zurich }

\msk\nin
M\o ller AP, 
Birkhead TR. 1992.
A pairwise comparative method as illustrated by copulation frequency 
in birds.  {\it Amer.\ Nat.\/}~{\bf 139}, 644--56. 

\msk\nin
Pagel MD. 1993.
Seeking the evolutionary regression coefficient: an analysis of what 
comparative methods measure.
{\it J.\ Theor.\ Biol.\/}~{\bf 164}, 191--205.
\msk

\msk\nin
Pagel MD, 
Harvey  PH. 1989.
Taxonomic differences in the scaling of brain size on body size among mammals. 
{\it Science\/}~{\bf 244}, 1589--93. 


\msk\nin
Purvis A, 
Rambaut A.  1994.
{\it Comparative analysis by independent contrasts (CAIC),\/} Version 2. 
Oxford University. 

\msk\nin
Scheff\'e H. 1959. 
{\it Analysis of variance.\/}  New York: Wiley. 

\msk\nin
Stearns SC. 1992. 
{\it The evolution of life histories.\/}
Oxford University Press. 

\msk\nin
Thompson GG, 
Withers PC. 1998.
Standard and maximal metabolic rates of goannas (Squamata: Varanidae). 
{\it Physiol.\ Zool.\/}~{\bf 70}, 307--23. 

\msk\nin
Younger MS. 1985.
{\it A first course in linear regression,\/} 2nd Edn. Boston: PWS Publishers.

\newpage
\section*{Appendix}

\subsection{The O-U model.}

The O-U model for the error structure can be constructed as follows.
We begin with a phylogeny consisting of a tree~$T$ with a root~$0$,
$n$ leaves and a set~$E$ of edges.  The length of an edge~$e$ is
denoted by~$\ell(e)$, and the distance from its rootward node to the 
root by~$t(e)$.  There is 
a unique path~$P_i = (e_{i1},e_{i2},\ldots,e_{im_i})$ from the root 
to each leaf~$i$ in~$T$, $1\le i\le n$, with
$0 = t(e_{i1}) < t(e_{i2}) < \cdots < t(e_{im_i})$, with
$t(e_{ij}) = t(e_{i,j-1}) + \ell(e_{i,j-1})$ for each $2\le j\le m_i$,
and with leaf~$i$ at distance $d_{0i} = t(e_{im_i}) + \ell(e_{im_i})$
from the root.

An $\OU(\t^2,\l)$--process~$X$ has the property that, for any $s,u>0$,
the conditional distribution of~$X(s+u)$ given $X(s)=x$ is that of
$xe^{-\l u} + X_0(u)$, where~$X_0$ is an $\OU(\t^2,\l)$--process
with $X_0(0) = 0$, and thus $X_0(u)$ is normally distributed with
mean~$0$ and variance $\s^2(1-e^{-2\l u})$, where we write
$$
  \s^2 := \t^2/(2\l)
$$
for the equilibrium variance.  Hence the O-U error
model on the tree can be constructed by associating with each
edge~$e$ an {\it independent\/} normally distributed random 
variable~$Z(e)$ with mean~$0$ and variance $\s^2(1-e^{-2\l \ell(e)})$,
and by then defining the error at leaf~$i$ to be
\eq\label{A1}
\e_i := Z\uo e^{-\l d_{0i}} 
   + \sjmi Z(e_{ij})e^{-\l(d_{0i}-t(e_{ij})-\ell(e_{ij}))}, 
\en
where~$Z\uo$ denotes the value of the error at the root.  In our
formulation, in which the O-U process stationary, $Z\uo$ is an independent
normal random variable with mean~$0$ and variance~$\s^2$.

It is now simple to deduce from~\Ref{A1} and from the independence
of the~$Z(e)$'s that $\ex \e_i = 0$ for all~$i$ and that (as has to be,
because of stationarity)
\eqs
\var\e_i &=& e^{-2\l d_{0i}}\var \{Z\uo\} 
   + \sjmi \var \{Z(e_{ij})\}e^{-2\l(d_{0i}-t(e_{ij})-\ell(e_{ij}))}\\
 &=& \s^2e^{-2\l d_{0i}}\Blb  1 +
   \sjmi  \{1 - e^{-2\l \ell(e_{ij})}\}e^{2\l(t(e_{ij})+\ell(e_{ij}))}\Brb\\ 
 &=& \s^2,
\ens
since the sum telescopes because $t(e_{ij}) = t(e_{i,j-1}) + \ell(e_{i,j-1})$,
and since $t(e_{i1}) = 0$ and $d_{0i} = t(e_{im_i}) + \ell(e_{im_i})$.
For the covariances, we similarly have
$$
\cov(\e_i,\e_l) = e^{-\l(d_{0i}+d_{0l})}\var\{Z\uo\}
  + \sjmij \var \{Z(e_{ij})\}
	   e^{-\l(d_{0i}+d_{0l}-2\{t(e_{ij})+\ell(e_{ij})\})},
$$
where 
$$
  P_i\cap P_l = (e_{i1},\ldots,e_{im_{il}}) = (e_{l1},\ldots,e_{lm_{il}})
$$
is the overlap between the paths leading from the root to $i$ and~$l$.  Hence
\eqa
\cov(\e_i,\e_l) &=& e^{-\l(d_{0i}+d_{0l})}\Blb 1 +
   \sjmij \{1 - e^{-2\l \ell(e_{ij})}\}e^{2\l(t(e_{ij})+\ell(e_{ij}))}\Brb\non\\
 &=& \s^2 e^{-\l\{(d_{0i}-t(e_{im_{il}})-\ell(e_{im_{il}}))
    + (d_{0l}-t(e_{lm_{il}})-\ell(e_{lm_{il}}))\}} \non\\
 &=& \s^2 e^{-\l (d_i + d_l)} = \s^2 e^{-\l d_{il}}, \label{A2}
\ena
where
$$
  d_i = d_{0i}-t(e_{im_{il}})-\ell(e_{im_{il}}) \quad{\rm and}\quad
	d_l = d_{0l}-t(e_{lm_{il}})-\ell(e_{lm_{il}}),
$$
and $d_{il}=d_i+d_l$ is the tree--distance from $i$ to~$l$, since
$e_{im_{il}} = e_{lm_{il}}$ and this is the last common edge in the
paths $P_i$ and~$P_j$. Equation~\Ref{(3)} now follows immediately. 

There is a non--trivial limit as $\l\to0$.  In this limit, each~$Z(e)$ is normally
distributed with mean~$0$ and variance~$\ell(e)\t^2$, so that the
joint {\it conditional\/} distribution of the~$\e_i$'s given any fixed
value~$z_0$ of~$Z\uo$ is the
same as for the Brownian model of evolution with infinitesimal 
variance~$\t^2$ and having value~$z_0$ at the
root.  Equivalently, one can check that the covariance structure of
the $\OU(\t^2,\l)$ model, restricted to the space of linear combinations of 
$\e_i-\bar\e$, $1\le i\le n$, converges to that of the Brownian model
on the same space, where $\bar\e := n^{-1}\sn\e_i$, irrespective
of the value of~$z_0$.  Note that this limiting model does {\it not\/}
have an equilibrium distribution.

   Instead of taking~$Z\uo$ to have the stationary distribution in
the O-U model, Blomberg \etal~(2003) suppose that $\var Z\uo = 0$, so 
that~$Z\uo$ is considered to be fixed at some (unspecified) value~$z_0$.
Thus their formulae for variances and covariances differ from those
above, in that the first term in each sum is lost, giving, in our
notation,
\eqs
\var'\e_i &=& \s^2(1-e^{-2\l d_{0i}});\\
\cov'(\e_i,\e_l) &=& \s^2(e^{-\l d_{il}} - e^{-\l(d_{0i}+d_{0l})}),
\ens
so that the root to leaf distances also enter their formulae.  To this
added complication comes the problem of the means; they now have
$\ex'(\e_i) = z_0 e^{-\l d_{0i}}$.  Thus the usual linear model analyses,
conducted on the assumption that errors have zero mean, are inconsistent
with their formulation (at least, if the $d_{0i}$'s are not all equal
and if $0 < \l < \infty$) unless~$z_0$ is fixed to be zero.
Hence their formulae are in general
only valid if a time earlier than the first split in the phylogeny
is known, {\it at which	the error is known to be exactly\/}~$0$, 	 
and if this time is then taken to be the root.  This seems to be 
rather an unlikely circumstance, and there is certainly no way of
inferring the value of such a time from an inter--species distance
matrix.  Hence the stationary O-U model is much to be preferred;
it pre-supposes merely that $\OU(\t^2,\l)$--style evolution had
already been taking place for a reasonable length of time before
the first split in the phylogeny.

Butler \& King~(2004), who are principally interested in more
detailed modelling of adaptive evolution, have a very ingenious
approach to~$Z\uo$. They treat its value as a {\it parameter\/}
of the model, to be estimated along with $\t^2$ and~$\l$.  This
approach again has the disadvantage of including more elements of the
phylogeny in the formulae.  It is also not clear that inference
about $\b\ui,\ldots,\b\uk$ would remain invariant using their
approach, if the root were moved further into the past from the time 
of the first split in the phylogeny, while leaving the rest of the
phylogeny unchanged; this should, however, logically be the case.
In view of these considerations, our simpler model would seem to
be preferable here also.

\subsection{Estimation as a function of~$\l$.}

To illustrate how the model estimated from data varies with the
choice of~$\l$, consider the case in which there are no covariates,
so that $y_i = \m + \e_i$.  Then the statistic
$$
  S^2 := {1\over n-1}\sn (y_i - \bar y)^2 = {1\over 2n(n-1)}
	  \sn\sln (y_i-y_l)^2,
$$
for a model with independent and identically distributed errors,
is a natural estimator of the common variance of the~$\e_i$'s.  For
the O-U error model, we have
\eq\label{A3}
  \ex\{(\e_i-\e_l)^2\} = 2\s^2(1 - e^{-\l d_{il}})
	  = {\t^2\over\l}(1 - e^{-\l d_{il}}), 
\en
from~\Ref{A2}, so that~$S^2$ estimates $\t^2 D_\l$, where
$$
  D_\l := {1\over 2n(n-1)}\sn\sln {1\over\l}(1 - e^{-\l d_{il}}).
$$
Hence, for given~$\l$, a reasonable (but in general not optimal)
estimator of~$\t^2$ is given by
\eq\label{A4}
  \hat\t^2_\l := S^2/D_\l.  
\en

If $\l$ is very large, then $D_\l \sim 1/(2\l)$, and it thus follows
from~\Ref{A4} that~$S^2$ estimates the equilibrium variance $\s^2 = \t^2/(2\l)$,
as is to be expected close to the model of independent errors.  Note,
however, that~$S^2$ is a fixed function of the data, so that, from~\Ref{A4},
the sequence of models estimated as~$\l$ increases has $\hat\t^2_\l \sim 2\l S^2$
growing to infinity linearly with~$\l$. Thus, in this limit, the rate
of random disturbances estimated from a fixed set of data tends to infinity.

If, on the other hand, $\l\to 0$, then $D_\l$ increases to its 
maximal value of 
$$
  D_0 = {1\over 2n(n-1)}\sn\sln d_{il},
$$
and~$\t^2$ is estimated by $S^2/D_0$, as appropriate for the Brownian model.

In the limit as $\l\to0$, the curiosity is rather the limiting
value of~$1$ for $\Corr(\e_i,\e_l)$, as implied by~\Ref{(3)}. The reason
for this is as follows.  Suppose that $X$, $V_1$ and~$V_2$ are independent
random variables, and that $X_1 := X + V_1$, $X_2 := X + V_2$.  Then
\eqs
  \Corr(X_1,X_2) &=& \var X \Big/ \sqrt{(\var X + \var V_1)(\var X + \var V_2)}\\
	  &=& 1/\sqrt{(1 + \h_1)(1 + \h_2)}, 
\ens
where $\h_j = \var V_j / \var X$ for $j=1,2$.  If now $\var V_1$ and~$\var V_2$
remain fixed, but $\var X \to \infty$, it follows that $\Corr(X_1,X_2) \to 1$;
and this despite the fact that
$$
  \ex\{(X_1-X_2)^2\} = \{\ex V_1 - \ex V_2\}^2 + \var V_1 + \var V_2
$$
remains constant.	
Comparing this setting with that of~\Ref{(4)}, it follows that the errors $\e_i$
and~$\e_l$ for species $i$ and~$l$ have correlation~$1$, in the limit as
$\l\to0$, only
because they inherit the same element~$X_0$ from their common ancestor species,
whose variance $\s^2 = \t^2/(2\l)$ tends to infinity as $\l\to0$.
In particular, as implied by~\Ref{A3}, $\ex\{(\e_i-\e_l)^2\}$ remains bounded
away from zero and infinity as $\l\to0$, so that the random variability 
between species does not disappear, even though $\Corr(\e_i,\e_j) \to 1$.
Indeed, the expression~\Ref{A3} for $\ex\{(\e_i-\e_l)^2\}$ provides
a consistent basis for comparing variability and strength of dependence
across different models, and is used as such in Section~3:
c.f.~also Math\'eron's~(1965) variogram.

\subsection{Departures from equilibrium.}

The analysis proposed in this paper supposes that the underlying O-U 
error process is in equilibrium.  This is not a problem unless, as
noted in Section~2, values of~$\l$ are considered which are so small
that it is doubtful whether equilibrium could ever have been reached.
However, the value $\l=0$ (for which there can be no equilibrium) represents
the well-tried Brownian model of evolution, and it is therefore
reasonable to ask how our procedure behaves for~$\l$ very close to~$0$.

Since we base our analysis only on the centred variables $y_i - \bar y$,
it is enough to understand what happens for {\it differences\/} of
errors $\e_i-\e_l$ between pairs of species.  So suppose that the root
value~$Z\uo$ in~\Ref{A1} is not at equilibrium, but instead has a normal
distribution with mean~$\m$ and variance~$v$ given by
$$
   \m := s\,\Bl{\t\over\sqrt{2\l}}\Br\,e^{-\l D} \quad {\rm and}\quad
	 v := \Bl{\t^2\over{2\l}}\Br\,(1 - e^{-2\l D});
$$
this represents an initial value for the error in the ancestor species
of~$s$ standard deviations away from zero, at an epoch~$D$ units of 
evolutionary time prior to the root.	 Then it is easy to calculate
\eqs
  \ex(\e_i-\e_l) &=& \ex\{Z\uo(e^{-\l d_{0i}} - e^{-\l d_{0l}})\} \\
	&=&  s\,\Bl{\t\over\sqrt{2\l}}\Br\,e^{-\l D}\, 
	  e^{-\l D_{il}}(1 - e^{-\l \d_{il}}),
\ens
where $D_{il} = \min\{d_{0i},d_{0l}\}$, $\d_{il} = |d_{0i}-d_{0l}| \le d_{il}$,
and $d_{0i}$ represents the evolutionary time from the root until species~$i$.
Hence, writing $\D_{il} := D + D_{il}$ for the time from the initial epoch
until the $i$ and~$l$ ancestries split, we have
\eq\label{A5}
  \ex(\e_i-\e_l) = s\,\Bl{\t\over\sqrt{2\l}}\Br\,e^{-\l \D_{il}}\, 
		(1 - e^{-\l \d_{il}}),  
\en
to be compared with the equilibrium standard deviation of $\e_i-\e_l$
which, from~\Ref{A3}, is given by
\eq\label{A6}
  {\rm SD}\,(i,l) := \Bl{\t\over\sqrt{\l}}\Br\,\sqrt{1 - e^{-\l \d_{il}}}.  
\en	  
The ratio of these two quantities is thus
$$
  {s\over2}\,	{e^{-\l \D_{il}}\,(1 - e^{-\l \d_{il}}) \over 
	  \sqrt{1 - e^{-\l d_{il}}}}	=: {s\over2}\,r(i,l;\l),
$$
say, where
$$
   r(i,l;\l) \le \{e^{-\l \D_{il}}\sqrt{\l \D_{il}}\} \,\sqrt{\d_{il}/\D_{il}}.
$$
Thus there is {\it no mean effect\/} if the branch lengths to $i$ and~$l$
are equal ($\d_{il}=0$), or if $\l=0$, or if $\l = \infty$, or if $s=0$; 
more generally,
the effect is small if $\l \D_{il}$ is either small or large, and the
largest effect possible, occurring when $\l = 1/(2\D_{il})$, is
${s\over 2\sqrt{2e}} \sqrt{\d_{il}/\D_{il}}$.  Hence, if the 
{\it differences\/} in branch lengths are much smaller than the
times from the initial epoch until species diverge, there can be no
appreciable mean effect.

For the variability, the conclusions are entirely analogous. It is
straightforward to compute
$$
  \var\{\e_i-\e_l\} - \{{\rm SD}\,(i,l)\}^2 \ =\ 
	  \Bl{\t^2\over{2\l}}\Br\,e^{-2\l \D_{il}}\,(1 - e^{-\l \d_{il}})^2,
$$
and the ratio of this quantity to $\{{\rm SD}\,(i,l)\}^2$, the relative
error in the variance induced by assuming the process to be in equilibrium, 
gives the value
$\half \{r(i,l;\l)\}^2$.  Once again, there is no correction to be made
if the branch lengths to $i$ and~$l$ are equal ($\d_{il}=0$), or if $\l=0$, 
or if $\l = \infty$; and the largest effect possible is
${1\over 4e}\{\d_{il}/\D_{il}\}$. Hence, if the 
{\it differences\/} in branch lengths	are much smaller than the
times from the initial epoch until species diverge, there can be no
appreciable effect on the variances, either.

\vfil\eject

\nin{\bf Table 1:} Testis size, mean HTL, duct length and spermathecal size
for 15 species of Scathophagidae: data from Minder \etal~(2005).

\halign{
#\hfil&\hfil#\hfil&\hfil#\hfil&\hfil#\hfil&\hfil#\hfil\cr
 Species  &\ Testis size \ &\ mean HTL \ &\  
Duct Length \  &\ Spermathecal Area \ \cr
 & (mm$^2$) & (mm) & (mm) & (mm$^2$) \cr
 & $y$ & $x\ui$ & \phantom{1}$x\ut$ & $x\uh$\cr	
{\it Cordilura albipes\/}	  &.169 &2.410	&\phantom{1}.534 &.00490 \cr
{\it Cleigastra apicalis\/}	  &.078	&2.080	&\phantom{1}.412	&.00769 \cr	
{\it Cordilura ciliata\/}         &.435	&3.290	&\phantom{1}.604	&.01743	\cr
{\it Cordilura pubera\/}	  &.332	&2.775	&\phantom{1}.727	&.01611	\cr
{\it Microprosopa pallidicauda\/} &.477	&2.125	&\phantom{1}.531	&.00795	\cr
{\it Norellia liturata\/}	  &.382	&2.095	&\phantom{1}.962	&.02497	\cr
{\it Norellia spinimana\/}	  &.547	&2.295	&1.086	&.02048	\cr
{\it Norellia striolata\/}	  &.855	&3.110	&1.384	&.02397	\cr
{\it Phrosia albilabris\/}	  &.319	&2.380	&\phantom{1}.519	&.01485	\cr
{\it Scathophaga cineraria\/}  	  &.486	&2.750	&\phantom{1}.561	&.01046	\cr
{\it Scathophaga furcata\/}	  &.965	&2.430	&\phantom{1}.541	&.02195	\cr
{\it Spaziphora hydromyzina\/}    &.134	&2.000	&\phantom{1}.235	&.01049	\cr
{\it Scathophaga stercoraria\/}	  &.544	&2.815	&\phantom{1}.672	&.01044	\cr
{\it Scathophaga suilla\/}	  &.461	&2.380	&\phantom{1}.386	&.01002	\cr
{\it Scathophaga taeniopa\/}	  &.699	&2.695	&\phantom{1}.479	&.01347	\cr
}
\vfil\eject
\nin{\bf Table 2:} Numbers of differing pairs between the sequences
 of 810 mDNA letters coding for the COI gene in 15 species of 
Scathophagidae: original sequences from Genbank.\medskip

\halign{
\hfil\hfil#&\hfil#&\hfil#&\hfil#&\hfil#&\hfil#&\hfil#&\hfil#&\hfil#&
\hfil#&\hfil#&\hfil#&\hfil#&\hfil#&\hfil#&\quad#&\ #\hfil\cr
 \phantom{XXXX} a & b & c & d & e & f & g & h & i & j &k &l &m &n &o &&Species \cr
  * &70 &64 &69 &72 &73 &87 &83 &62 &67 &75 &82 &63 &65 &70 &a: &{\it Cordilura albipes\/}\cr
    & * &92 &93 &64 &69 &76 &75 &81 &60 &71 &74 &55 &54 &61 &b: &{\it Cleigastra apicalis\/}\cr
    &   & * &63 &80 &77 &88 &88 &71 &73 &68 &89 &71 &72 &77 &c: &{\it Cordilura ciliata\/}\cr
    &   &   & * &81 &83 &99 &96 &75 &84 &88 &96 &81 &81 &85 &d: &{\it Cordilura pubera\/}\cr
    &   &   &   & * &66 &79 &73 &82 &63 &74 &55 &52 &59 &63 &e: &{\it Microprosopa pallidicauda\/}\cr
    &   &   &   &   & * &67 &65 &77 &59 &70 &70 &58 &57 &63 &f: &{\it Norellia liturata\/}\cr
    &   &   &   &   &   & * & 9 &95 &67 &80 &74 &69 &72 &79 &g: &{\it Norellia spinimana\/}\cr
    &   &   &   &   &   &   & * &94 &64 &77 &71 &64 &68 &75 &h: &{\it Norellia striolata\/}\cr
    &   &   &   &   &   &   &   & * &69 &78 &83 &68 &70 &72 &i: &{\it Phrosia albilabris\/}\cr
    &   &   &   &   &   &   &   &   & * &36 &62 &35 &20 &30 &j: &{\it Scathophaga cineraria\/}\cr
    &   &   &   &   &   &   &   &   &   & * &73 &43 &41 &45 &k: &{\it Scathophaga furcata\/}\cr
    &   &   &   &   &   &   &   &   &   &   & * &56 &56 &62 &l: &{\it Spaziphora hydromyzina\/}\cr
    &   &   &   &   &   &   &   &   &   &   &   & * &29 &38 &m: &{\it Scathophaga stercoraria\/}\cr
    &   &   &   &   &   &   &   &   &   &   &   &   & * &22 &n: &{\it Scathophaga suilla\/}\cr
    &   &   &   &   &   &   &   &   &   &   &   &   &   & * &o: &{\it Scathophaga taeniopa\/}\cr
}

\end{document}